\RequirePackage{fix-cm}
\documentclass{svjour3}                     
\smartqed  
\usepackage{graphicx}
\usepackage{amsmath}
\usepackage[numbers]{natbib}
\usepackage{hyperref}
\usepackage{xurl}
\usepackage{caption}
\usepackage{rotating}
\usepackage{lscape}    
\usepackage{multirow, makecell, caption}
\usepackage{soul}
\usepackage{xcolor}

\journalname{Applied Intelligence}
\begin{document}

\title{TaDeR: A New Task Dependency Recommendation for Project Management Platform
}

\titlerunning{TaDeR: A New Task Dependency Recommendation}        

\author{Quynh Nguyen \and
        Dac H. Nguyen  \and
        Son T. Huynh \and
        Hoa K. Dam \and
        Binh T. Nguyen
}

\institute{Quynh T.N.Nguyen \at
              AISIA Research Lab, Ho Chi Minh City, Vietnam\\
           \and
           Dac H. Nguyen \at
              AISIA Research Lab, Ho Chi Minh City, Vietnam\\
            \and
           Son T. Huynh \at
              AISIA Research Lab, Ho Chi Minh City, Vietnam\\
            \and
           Hoa Khanh Dam \at
              University of Wollongong, Australia\\
              \and
              Binh T. Nguyen (Corresponding Author) \at
              Vietnam National University in Ho Chi Minh City\\
              University of Science, Vietnam\\
              \email{ngtbinh@hcmus.edu.vn}}

\date{Received: date / Accepted: date}

\maketitle

\begin{abstract}

Many startups and companies worldwide have been using project management software and tools to monitor, track and manage their projects. For software projects, the number of tasks from the beginning to the end is quite a large number that sometimes takes a lot of time and effort to search and link the current task to a group of previous ones for further references. This paper proposes an efficient task dependency recommendation algorithm to suggest tasks dependent on a given task that the user has just created. We present an efficient feature engineering step and construct a deep neural network to this aim. We performed extensive experiments on two different large projects (MDLSITE from moodle.org and FLUME from apache.org) to find the best features in 28 combinations of features and the best performance model in the combination of two embedding methods (GloVe and FastText). We consider three types of models (GRU, CNN, LSTM) using Accuracy@K, MRR@K, and Recall@K (where  K = 1, 2, 3, and 5) and baseline models using traditional methods: TF-IDF with various matching score calculating such as cosine similarity, Euclidean distance, Manhattan distance, and Chebyshev distance. After many experiments, GloVe Embedding and CNN model reached the best result in our dataset so we decided to choose this model as our proposed method. In addition, adding the time filter in the post-processing step can significantly improve the recommendation system's performance. The experimental results show that our proposed method can reach 0.2335 in Accuracy@1 and MRR@1 and 0.2011 in Recall@1 of dataset FLUME. With the MDLSITE dataset, we obtained 0.1258 in Accuracy@1 and MRR@1 and 0.1141 in Recall@1. In the top 5, our model reached 0.3040 in Accuracy@5, 0.2563 MRR@5, and 0.2651 Recall@5 in FLUME. In the MDLSITE dataset, our model got 0.5270 Accuracy@5, 0.2689 MRR@5, and 0.2651 Recall@5. 



\end{abstract}

\section{Introduction}
\label{intro}
With the rapid development of science and technology, an increasing number of startups and companies are providing many products related to software engineering, information technology, business, operation, and more. Each company can have multiple projects for different products every year. It is crucial to have a better way of managing human resources, task planning, and ensuring that projects can meet deadlines and milestones.

There are various project management platforms in the market. Each company can easily manage necessary features in every project, including user stories and issues, plan sprints, task assignments, team collaboration, real-time reporting, resource management, billing, role management, milestones, and deadlines. Several well-known project management platforms are Asana\footnote{https://asana.com/}, Atlassian Jira\footnote{https://www.atlassian.com/software/jira}, Trello\footnote{https://trello.com/en}, Zoho Projects\footnote{https://www.zoho.com/projects/}, and BaseCamp\footnote{https://basecamp.com/}. These platforms have different types of features related to project management. There are available options to choose a free version or purchase a business one with better support and functionalities. 

Using a project management tool would help teams alleviate problems such as lack of communication among teammates, lack of alignment across sub-teams, the difficulty of keeping clients and partners on board, or the time limit constraint of chosen projects. It is also essential to consider the financial ability to use project management tools. However, whenever it is set up and gets running, project management software can bring more value to the team's projects by creating the central hub of socializing and exchanging information to account for what everyone is doing. In addition, it can help project stakeholders monitor a project's progress. and enable them to see what they could do better to get the project moving. 

There are multiple valuable features integrated into a project management software. Using a project management software, one can have all they need to manage projects at their disposal. Users can quickly buzz each other for support in individual or group chat windows, share their work on standard progress pages, and exchange multimedia data efficiently like they do in private communications. Project management systems can automatically save the history of all communications in the team and organize it to make it easy to retrieve old data at any time we need. One of the most useful features is the real-time notifications on all project changes so that everyone in the team can quickly notice and avoid any missing issues later.

Although existing project management tools are useful, they lack advanced analytical methods that are capable of harvesting valuable insights from project data for prediction, estimation, planning and action recommendation. Many decision-making tasks in projects are still performed by teams without machinery support. 

To provide such support, we propose a \underline{Ta}sk \underline{De}pendency \underline{R}ecommendation system, namely TaDeR\footnote{Tader is a small river in Spain.} that can be integrated into a project management tool.  TaDeR recommend interdependencies between a newly created tasks and existing tasks using a range of machine learning techniques. Using information in the newly created task, TaDer is able to suggest to the user the top $K$ tasks that can link to this task. There are several steps to construct this system using a machine learning approach. First, we formulate the problem as a recommendation system using historical data from a given project. Then, we collect multiple data sources to investigate the main problem and utilize text mining techniques to preprocess the input data, extract useful features for the problem, and construct the most suitable model for each data source. Finally, we propose one approach for the problem using the Siamese architecture and CNN for feature extraction and model training. 

Our TaDeR system can provide a complete pipeline from data collection, data processing, model training, and model prediction. We consider three different performance metrics for choosing the best model for each dataset, including Accuracy@K, MRR@K, and Recall@K.($K=1, 2, 3, 5$). 

In summary, our main contribution can be described as follows:
\begin{enumerate}
 \item Very few previous studies are related to the task dependency recommendation system for project management platforms. To the best of our knowledge, our work is the first of such a study that can directly contribute to the problem and demonstrate results in both model training and framework.
 
\item Through multiple experiments with different combinations of useful features, we demonstrate that using the Siamese architecture \cite{siamese} for constructing the corresponding recommendation algorithm produces promising results. The proposed method can obtain accuracy higher than 0.3 from top 5 in all benchmark datasets.

\item Applying the ``time filter'' to the main problem can improve the TaDeR system's performance. The experimental results show that most of the tasks created during the last several months have a high chance of linking to the current one. 

\item We publish all codes and datasets related later to contribute our work to the research community-related, available at \href{https://github.com/thuynguyetquynh/TaDeR-A-New-Task-Dependency-Recommendation-for-Project-Management-Platform}{a link}.
Since the data is quite large, we shared it at \href{https://www.dropbox.com/sh/v9foat8opqyl88y/AAD-NzagFLIHISihwyW6XPjpa?dl=0}{a link}.
\end{enumerate}

The structure of the paper is as follows. We provide an overview and the relevant background of our TaDeR system in Section 3. We describe our approach, including data processing, feature extraction, and model training in Section 4. After that, we illustrate our evaluation step in Sections 5. All experimental results are illustrated in Section 6, and finally, we give our conclusion and future work in the last section.

\section{Related Work}
There have been various works related to issue-link detection for Jira tasks. Choetkiertikul and colleagues \cite{Hoa_2015} investigated the delaying time prediction problem in software projects and proposed a novel method to estimate the risk of being delayed for ongoing software tasks. This approach helps project managers and decision-makers proactively determine all potentially risky tasks and optimize the overall costs (including the cost of human resources and infrastructure ones). The paper also introduced a proper feature engineering step by utilizing the existent factors of individual software tasks and other features related to these features' interaction. The experimental results showed that the proposed method could outperform the previous approaches in precision, recall, F1-score, and AUC. 

 Lee et al. \cite{Lee_2017} studied a mechanical bug triage problem in the bug resolution process by considering a deep learning technique. They adopted word embedding techniques and Convolutional Neural Networks to construct appropriate features and a prediction model. The experiments revealed promising results with various applications in both industrial and open-source projects. 

Lankan and co-workers \cite{Lamkanfi_2010} introduced an interesting predictive approach to estimate the severity of a reported bug in software development projects. By extracting useful features from a reported bug's textual description, they constructed the corresponding classifier for severity prediction. In experiments, the paper compared the proposed technique in three different datasets from the open-source community (Mozilla, Eclipse, and GNOME) and obtained good performance in precision, recall, and F1-score. 

Pandey et al. \cite{Pandey_2017} investigated the mechanical classification problem for software issue reports by considering different machine learning techniques. This problem has many applications in reviewing reports submitted from software developers, testers, and customers and optimizing the development time for each software project. The paper considered various classifiers (including naive Bayes, K-nearest neighbors, linear discriminant analysis (LDA), SVMs with different kernels, decision trees, and random forests) and compared their performance in three open-source projects. The experimental results showed that random forests outperformed the remaining classification methods in accuracy and F-1 scores. 

Lam and colleagues \cite{Lam_2017} presented a fascinating approach for automatically detecting all potential coding files having bugs in software projects for a given bug report. By combining deep learning features, information retrieval (IR) techniques, and projects' bug-fixing historical data, the authors indicated the proposed algorithm's better performance than previous state-of-the-art IR and machine learning techniques.
One can find other works related to bug reports at \cite{Tian_2015, Xia_2017, YAN201637}. 

Runeson and colleagues \cite{Runeson_2007} investigated the duplicate detection for defect reports using various natural language processing (NLP) techniques and obtained an accuracy of about $66.67\%$ when analyzing defect reports at Sony Ericsson Mobile Communications. 
Other studies related to the duplicate detection for bug reports are detailed at \cite{ Wang_2008, Sun_2011}.

\section{Methodology}

\subsection{Background}


\subsubsection{LSTM \& GRU}


As one of the most powerful and well-known neural networks, Long Short-Term Memory was initially proposed by Hochreiter and Schmidhuber \cite{LSTM}. This neural network may be considered an upgrade version of the Recurrent Neural Network (RNN) \cite{RNN}. Least squares time-series modeling (LSTM) is an artificial neural network designed to identify patterns in sequences of data such as a sentence, document, and numerical time series data. By utilizing an RNN-based layer, such as the LSTM, while analyzing a phrase, we can intuitively depict the influence of surrounding words on the current processing word, which is useful in many situations. The output of LSTM can be differentiated in this manner by using the same processing word but in a different location in a phrase or with other surrounding words that are different.

In particular, because of RNN's inherent ability, LSTM "remembers" long-term or short-term reliance, which implies that the efficacy of a word seems to be diminished when it is located far away from the processing word and vice versa. In mathematical formulae, it may be represented as a three-gate structure, which includes the input gate, the forget gate, and the output gate, among other things. The Forget gate determines whether information from the previous cell should be kept and which information should be deleted. The input gate determines what information should be obtained from the input and concealed state from the previous state by analyzing the prior state. The output gate determines the output from this cell state to determine the output from the following concealed state.

GRU is a technique developed by Cho et al. \cite{GRU} to address the gradient vanishing problem that occurs while using a recurrent neural network. Because they are built similarly, GRU is considered a variation of LSTM. In certain instances, the outcomes may be just as favorable. GRU is comprised of two gates. Update gate has a role to decide how much past information to forget, and Reset gate role to decide what information to throw away and what new information to add.

GRU has fewer parameters than LSTM's as it does not have the output gate. One of the main differences between a normal RNN and the GRU is the stealth control, which allows us to learn mechanisms to decide when to update and clear hidden states. 


\subsubsection{CNN}

Convolutional neural networks (CNNs) have emerged in the broader field of deep learning in the last few years, with unprecedented results across a variety of application domains, including image and video recognition, recommendation systems, image classification, medical image analysis, natural language processing, and financial time series analysis. In many cutting-edge deep neural network topologies, CNNs play a critical role. Conv 1D or 1D CNN is used as a feature extractor in this work after embedding all strings from the input.

The characteristics of an image may be extracted using a typical CNN. A picture and some sort of filter are the first two inputs that CNN takes into consideration (or kernel). This neural network (CNN) only examines a tiny portion of input data, and it shares parameters with all neurons to its left and right (since these numbers all result from applying the same filter). Until there is no longer a filter, this cycle will be repeated indefinitely. The input is in 2D rather than 3D when using Conv 1D.

Let the input x to convolution layer of length n and let the kernel h of size z. Let the kernel window be shifted s positions (number of strides) after each convolution operation, and p is the number of window kernels. Then, the convolution between x and h for stride s can be defined as:

\begin{equation}
  y(n) =
    \begin{cases}
      \sum^z_{i=0}x(n+1)h(i) & \text{, if } n=0  \\
      \sum^z_{i=0}x(n+i+(s-1)h(i) & \text{, otherwise}\\
    \end{cases}       
\end{equation}

\subsection{Word Embedding}
Most of the features in this project are textual features. Hence, we need a method to convert these features to number vectors, a learned form, to process them through the models. 
There are many popular embedding methods such as TF-IDF, Word2vec, GloVe, FastText, etc.

This paper used TF-IDF as a traditional method to compare with GloVe and FastText in the neural network approach. 

\paragraph{TF-IDF} or Term Frequency-Inverse Document Frequency \cite{ref1} is a popular method that has a similar rule with Count Vector when it also focuses on the frequency of a word. However, TF-IDF also cares about the frequency of that word in the whole dataset, besides the corresponding frequency in each document.
    
Normally, Term Frequency (denoted as TF) is the number of times words appear in the text, which can computed as follows:
\begin{equation*}
  tf(w,d)=\frac{f(w,d)}{\max(\{f(w,d): w \in d\})},
\end{equation*}
where $tf(w,d)$ is the term frequency of the word $w$ in the document $d$, $f(w,d)$ is the number of the word $w$ existing in the document $d$, $\max(f(w,d): w \in d)$ is the number of occurrences of the word $w$ with the most occurrences in the document $d$.

Meanwhile, Inverse Document Frequency (denoted as IDF) can measure the importance of a word as:
\begin{equation*}
  idf(w,D)=\log\left({\frac{|D|}{|\{d \in D: w \in d)\}|}}\right),
\end{equation*}
where $idf(w,D)$ is the value IDF of the word $w$ in all documents, $D$ is the set of documents, and $|{d \in D: w \in d)}|$ is the number of documents including the word $w$ in $D$.
    
Finally, the TF-IDF value of a word in a text can be formulated as:
\begin{equation*}
  \text{TF-IDF}(w, d, D) = tf(w, d) \times idf(w, D).
\end{equation*}

\paragraph{GloVe}
GloVe (Global Vectors) \cite{pennington2014glove} is one of the new methods to construct word vectors (introduced in 2014). It is essentially built on top of the Co-occurrence Matrix. 
GloVe based on the idea: The semantic similarity between two words $i$, $j$ can be determined through the semantic similarity between word $k$ and each word $i$, $j$, the $k$ words with good semantic determinism are the words that make the ratio $\frac{P(k|i)}{P(k|j)}$ go to 1 or just 0. For example, if $i$ is ``cake", j is ``milk" and k is ``biscuit", then the ratio will be quite large since ``biscuit'' means closer to ``cake'' rather than ``milk'', otherwise, if we substitute k for ``computer'', (1) will be approximately equal to 1 because ``computer'' has almost nothing to do with ``cake'' and ``milk''.

\paragraph{FastText}
A major drawback of word2vec is that it can only use words in the dataset. To overcome this, we have FastText \cite{bojanowski2016enriching} which is an extension of Word2Vec, built by the Facebook research team in 2016. Instead of training for word units, it divides the text into small chunks called n-grams for the word. 

For example, the word ``vision" would be ``vis", ``isi", ``sio", and ``ion", the vector of the word apple would be the sum of all these. Therefore, it handles very well for cases when the word is not found in the vocabulary list.

\subsection{Distance Metrics}
In traditional approach, we used TF-IDF to get numeric vectors from textual features then computed matching score by several distance metrics: cosine similarity, euclidean distance, manhattan distance and chebysev distance.

\paragraph{Manhattan distance}
Manhattan distance seems to works well with high-dimensional data, in this case our embedding vectors. It computes distance between two vectors if they could only move right angles with no diagonal movement.
\begin{equation*}
    \textbf{D(x,y)} = \displaystyle\sum_{i=1}^{k} |x_i-y_i|
\end{equation*}

\paragraph{Chebysev distance}
It is defined as the maximum distance value along one axis.
\begin{equation*}
    \textbf{D(x,y)} = max_i|x_i-y_i|
\end{equation*}

\paragraph{Cosine similarity}
Cosine similarity is a metric to calculate matching score between two non-zero vectors. In this case, those vectors are our embedding textual vectors. This metric is computing as
\begin{equation*}
    \textbf{\textit{similarity}} = \frac{AB}{||A||||B||} = \frac{\displaystyle\sum_{i=1}^{n} A_iB_i} {\sqrt{\displaystyle\sum_{i=1}^{n} A^2_i}\sqrt{\displaystyle\sum_{i=1}^{n} B^2_i}}
\end{equation*}
where $A$ and $B$ are our vectors needed to be compared.

\paragraph{Euclidean distance}
Comparable with cosine similarity, there is another popular metric called euclidean Distance which is using widely in many machine learning applications. This metric is displayed as the equation.
\begin{equation*}
    \textbf{\textit{d(A,B)}} =  \sqrt{\displaystyle\sum_{i=1}^{n} (q_i-p_i)^2} 
\end{equation*}
where $A$ and $B$ are our textual vectors.

\section{Our Approach}

\subsection{Data Processing}

Data processing is essential in every natural language processing problem to remove all possible noises, punctuation marks, special characters and enhance proposed techniques' performance. In this work, we apply the following data processing steps.

\subsubsection{Textual Features}

\begin{figure*}[!htpb]
  \centering
  \includegraphics[width = 0.81\linewidth]{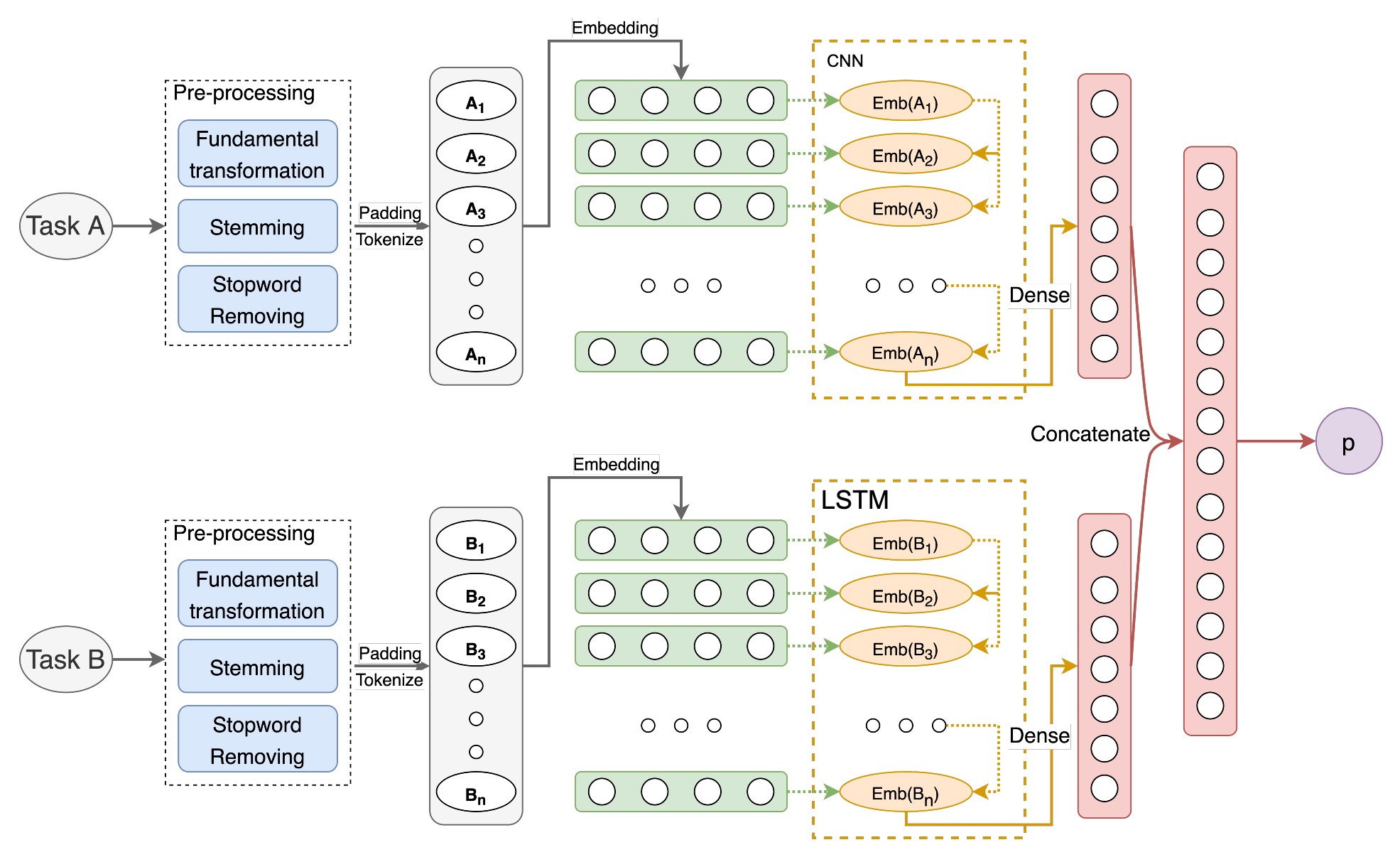}
  \caption{The architecture of our Siamese model: a CNN layer with the number of units of 256, and the Dense layer (Fully Connected layer) with the unit number of 256 using ReLU as the activation function, and the last layer using the activation function Softmax to compute the final matching score between two input Jira issues.} 
  \label{fig:siamese-model}
\end{figure*}

Removing noises and transforming words into nuggets that make the model extract value patterns are the data processing step's expectations. For Jira observations, textual features include various word forms, numbers, and noises, such as different words due to writing style, HTML's URL and elements, and stop words. We break those things down by solving each step-by-step in a suitable order, as shown in Figure 
\ref{fig:siamese-model}. 

Verbally, we first convert all the text in lower case and remove all the links, punctuation, numbers and stop words (the stop words list based on NLTK (Natural language processing toolkits proposed by Loper and his colleagues\cite{loper})'s stop words list\footnote{https://www.nltk.org/} which is combined with some words like ``e.g'', ``i.e'', ``http'',  ``htt'',  ``or'', and "www") from getting output as word-only features. Those "unstructured" features are passed into a transformative preprocess (applying stemming technique\footnote{https://nlp.stanford.edu/IR-book/html/htmledition/stemming-and-lemmatization-1.html} which is provided by NLTK) to convert them into their ``root forms''. Table \ref{tab:pre-processing} shows examples of data after those steps.

\begin{table}[]
\caption{Preprocessing results}
\centering
\label{tab:pre-processing}
\scalebox{0.93}{
\begin{tabular}{|l|l|l|}
\hline
\multicolumn{1}{|c|}{Datasets} & \multicolumn{1}{c|}{Original Text}                                                                                                                                                                                                                                                                                                                                                                                                                                                                                                       & \multicolumn{1}{c|}{Proccessed Text}                                                                                                                                                                                                                                                                                                                            \\ \hline
FLUME                          & \begin{tabular}[c]{@{}l@{}}$<p>$ RPM install runs as user flume\\ , but a file such as/var/log/messages \\ is default perms 600 on centos/redhat.\\  $</p>$ $<p>$Ideally we don’t want to run \\ flume node as root.$</p>$Flume node’s \\ tail source does not report error or go \\ into error state to user attempts to tail \\ a file it doesn’t have permissions to \\ read. Flume node’s tail source does \\ not report error or go into error state \\ to user attemptsto tail a file it doesn’t \\ have permissions to read.\end{tabular} & \begin{tabular}[c]{@{}l@{}}rpm install run user flume file var \\ log message default perm centos \\ redhat ideally want run flume \\ node root flume node tail source \\ report error go error state user \\ attempt tail file permission read \\ flume node tail source report error \\ go error state user attempt tail file \\ permission read\end{tabular} \\ \hline
MDLSITE                        & \begin{tabular}[c]{@{}l@{}}$</p>$Rather than writing docs AT \\ moodle DOT org, it would be nice to \\ have mail to links.$</p>$Add email \\ obfuscation to Moodle Docs Add \\ email obfuscation to Moodle Docs \\ Fix.\end{tabular}                                                                                                                                                                                                                                                                                                         & \begin{tabular}[c]{@{}l@{}}rather writing doc moodle dot \\ would nice mail to link add email \\ obfuscation moodle doc add email \\ obfuscation moodle doc\end{tabular}                                                                                                                                                                                        \\ \hline
\end{tabular}}
\end{table}

\paragraph{Feature Selection}

Among these 32 attributes collected \ref{tab:feature-description1}, \ref{tab:feature-description2}, \ref{tab:list-full-feature}, there is a huge number of these attributes have "NaN" values for most of datasets, including \emph{Labels, TimeOriginalEstimate, TimeEstimate, AggregateTimeOriginalEstimate, AggregateTimeRemainingEstimate, TimeSpent, and AggreateTimespent}.
There are only nine attributes available to fill in when one Jira task is created: ``title'', ``description'', ``environment'', ``summary'', ``type'', ``priority'', ``status'', ``reporter'', and ``component''. 

After analyzing the datasets, we discover that only three attributes (\emph{Title, Description, and Summary}) can contain the fundamental information of a Jira issue, such as, e.g., "What type of this issue is?", "How does it affect the product and other functions?" and "Why does it occur?" As a result, we decide to use these three attributes to construct a suitable recommendation model for the problem. 

Moreover, to enhance the performance of those features, we also tried to extract time features included \emph{"Created Date"} and \emph{"Updated Date"}. If they are considered independently, these values seem to be similar in a large group of issues. Hence, we combine them to two features called \emph{"Cre-cre"} and \emph{"Cre-up"}. They are following an assumption that if two issues are created near each other created day, or one issue is created near the day one other issue is updated, they tend to have links between each other.

We use the Pandas library to convert all time features from string to DateTime type for this feature. We then get the absolute value of the subtraction between two created dates or between the created date and the updated date, as represented above with each pair of issues. 

Assume that X is an issue that we need to recommend a list of relevant matters that link with it. Y is the issue that we aim to check if it likely has a link with X or not. In the feature engineering step, we add two features: the gap between the created dates of both X and Y and the gap between the created date of X and the updated date of Y (in short, we named it $CC$ and $CU$, respectively).

\subsection{Model Training}

This section aims to present two proposed approaches to the main problem. First, as the primary problem can be formulated as a classification problem, we want to estimate the matching score between the current task created and other previous ones to choose the top $K$ relevant items for the output results of the TaDeR system. To construct an appropriate recommendation model from each training dataset, we can split all Jira observations into two groups. One group contains all existent pairs of Jira issues that link with each other (Jira users created these links during the related project). Another group has all remaining couples of Jira issues having no connection with others, namely ``lonely issues". We can create initial datasets for training and testing the expected recommendation model by this approach. We aim to build a binary classification model for estimating the probability of having a connection between two given Jira observations using these two groups. We label each pair as ``1'' if the two issues have a link between them, and ``0'' if they are not relevant. 

However, we need to assume that we can only utilize available attributes when creating the current Jira issue. In the modeling process, we consider two different directions. In this first direction, we compute TF-IDF features for all text attributes in this first direction and combine them for the final feature vector for a given Jira issue. We estimate their cosine similarity distance based on the feature vectors extracted from two given Jira issues. However, the result is not so well even when we tried to replace cosine similarity with euclidean distance, manhattan distance, and Chebyshev distance. In the second approach, we construct a Siamese neural network, where we use GloVe as our embedding and CNN layer to compute the deep-learning features from each Jira issue's input data. After that, we calculate the similarity between two feature vectors computed by one Dense layer for getting the matching score of these two issues. We later use embedding methods with FastText and replaced CNN with GRU and LSTM.

For returning the list of recommended items, we calculate the matching scores between the chosen Jira issues and all other previous ones and determine the list of top $K$ relevant items. In what follows, we will briefly describe our proposed models.

\subsubsection{Traditional Methods}

In this work, we compare many methods (from traditional to deep learning) to get our proposed technique. 

In the traditional approach, we first started with TF-IDF to vectorize all essential features, including the title, the description, and the summary, for a given Jira issue. Then, after extracting the TF-IDF features from these three attributes, we combine them into a feature vector and use this type of feature vector to estimate the matching score of two given Jira issues using cosine similarity. 

In addition, instead of cosine similarity, we use Euclidean distance, Manhattan distance, and Chebyshev distance.

However, cosine similarity and Euclidean distance gave the best results on both datasets. 

One can find the corresponding experiment results related to this traditional approach in Table \ref{tab:flume_traditional_1}, Table \ref{tab:flume_traditional_2}, Table \ref{tab:mdlsite_traditional_1}, and Table \ref{tab:mdlsite_traditional_2}.

\subsubsection{Siamese Model}

Siamese \cite{siamese} is the model released by Yann Lecun et al., a variant of neural networks that allows us to compare the similarities and differences in a pair of two objects. Then, depending on the dataset provided, one can find a suitable technique to compute each object's feature vector via a shared neural network and estimate the final matching score at the end. 

The Siamese architecture used in Jira is built, as shown in Figure \ref{fig:siamese-model}. For two input Jira issues, we apply the necessary data processing step for each, do the padding tokenize step, and then feed them through an embedding layer to transform words into the same higher-dimensional feature space where all terms are more closely related. We employ a CNN layer to control how much of the previous state is retained and how many parts of the new state are the same as the primordial state. After this step, we can obtain two feature vectors for two initial Jira issues. Using a Dense layer with the Softmax activation function, the model can output the probability of linking these two input issues. Finally, we use the MSE as a loss function to update weights by backpropagation in the training process.

After trying combinations of CNN, GRU, LSTM with different embedding methods included  FastText and GloVe, and we found out that GloVe with CNN brings the best results.

There are many reasons for this:
\begin{itemize}
    \item CNN is able to extract well both local and position-invariant features. Our dataset, however including mostly free-text and discrete information technology major words as we can see in \ref{tab:pre-processing} so LSTM and GRU is not worked as well as CNN.
    \item Similar to the difference between Glove and Fasttext, Glove focus more on global statistics hence Glove's word vectors are more discrete in the space which lead to a better result.
\end{itemize}

\section{Data Collection and Evaluation}


Nowadays, Jira has become a proprietary issue-tracking product developed by Atlassian, a day-to-day bug tracking and agile management dashboard of many popular products and projects of Atlassian and other technology teams. It contains several essential pieces of information that are adequate to explore, such as title, description, a summary of an observation (each observation here is a task, a story, or a project), and their linkage information.

To investigate the task dependency recommendation problem, we find available public projects with over 3000 observations with at least 100 linkages among Jira observations. There are two well-known data sources, including Apache\footnote{https://projects.apache.org/projects.html} and Moodle\footnote{https://tracker.moodle.org/projects}. Both Apache's teams and Moodle's teams usually use Jira as their primary project management platforms. 

In experiments, we choose two large datasets for comparing the performance of different methods: one Moodle project (MDLSITE\footnote{(Moodle Community Sites) https://tracker.moodle.org/}) and Apache project FLUME\footnote{https://flume.apache.org/}). Then, we did experiments with 28 combinations of features on FLUME since it has fewer issues than MDLSITE. Hence the experiments will run faster. After getting the best combination of features, we test different models on FLUME and MDLSITE to find which model has the best performance.

\subsection{Data Collection}

The Apache software foundation (or denoted as Apache\footnote{https://www.apache.org/}) has multiple projects/products using Jira as their primary project management dashboard. Similarly, Moodle also has various projects/products using Jira, such as Moodle App (MOBILE), Moodle QA (MDLQA), and Moodle Community Sites (MDLSITE). These are public projects for issue tracking, and one can access these projects by their public APIs to quickly collect or search relevant Jira observations. 

Among the datasets, each Jira task collected has 32 attributes. We list these attributes and the corresponding description at Table \ref{tab:feature-description1}, Table \ref{tab:feature-description2}, and Table \ref{tab:list-full-feature}.

\subsection{Datasets}
This section briefly describes datasets, including MDLSITE and FLUME. In experiments, all data sets are split into two mutually exclusive groups: one training set and one testing set with a ratio of 3:1. Instead of randomly dividing them, we choose a date as a splitting point, which means all observations created before this splitting point are training set. Other tasks made after this date belong to the testing set. This division can help us mimic a practical application of the TaDeR system when we utilize historical data for recommending the relevant tasks for one Jira observation created.

Specifically, we can construct six corresponding datasets for datasets collected, including three training sets and three testing sets as follows. For the MDLSITE dataset, we choose the splitting point as the 4100th day, which means all observations created in the first 4100 days belong to the training set, and the remaining ones are in the testing dataset. Also, we select the splitting point for the FLUME dataset as the 1577th day.

\begin{table}[ht]
\centering
\caption{The details of datasets chosen FLUME and MDLSITE}
\label{tab:data-desc}
\begin{tabular}{|l|c|c|c|}
\hline
\multicolumn{1}{|c|}{}                           & \textbf{FLUME} & \textbf{MDLSITE} \\ \hline
\textbf{Issues}                                  & 3373            & 5910   \\ \hline
\textbf{\# Links}                                & 1664            & 4566    \\ \hline
\textbf{\# Training   Size}                      & 2502            & 4389    \\ \hline
\textbf{\# Available   Links (Training Dataset)} & 1266            & 3084    \\ \hline
\textbf{\# Testing Size}                         & 871             & 1521    \\ \hline
\textbf{\# Available   Links (Testing Dataset)}  & 398             & 1482     \\ \hline
\end{tabular}
\end{table}

Ordinarily, the contents in the Summary column are replicates of the Title column. For this reason, when we analyze these datasets, we discover that they have pretty similar contents, and the histograms of the number of words look a bit close.

\subsection{Performance Metrics}

We measure each method's performance using the following standard metrics in the recommendation system: Accuracy@K, Recall@K, and MRR@K.

\subsubsection{Accuracy@K}

For a given list of K recommended items, the metric Accuracy@K can be computed as follows:
\begin{equation*}
  Accuracy@K=\frac{TP@K+TN@K}{TP@K+TN@K+FP@K+FN@K}
\end{equation*}
In the top K recommended results, TP@K is the number of actual relevant pairs predicted to be related, TN@K is the number of pairs that are irrelevant and predicted to be irrelevant. FP@K is the number of pairs that are irrelevant but predicted to be relevant, and FN@k is the number of pairs that are relevant but predicted to be irrelevant.

\subsubsection{Recall@K}
Recall@K is one of the essential metrics that can be determined as:
\begin{equation*}
  Recall@K=\frac{TP@K}{TP@K+FN@K}
\end{equation*}

\subsubsection{Mean Reciprocal Rank@K}
The formula of Mean Reciprocal Rank@K (MRR@K) can be given as follows
\begin{equation*}
  MRR=\frac{1}{|K|}\sum_{i=1}^{|K|}\frac{1}{rank_i}, 
\end{equation*}
where $rank_i$ denotes the rank of the first relevant result and $K$ is the top relevant issues.

\section{Experiments}

There are three fascinating questions we aim to investigate related to the task dependency recommendation system in our experiments. First, what types of existing attributes from a new Jira task created are vital to the corresponding model? In practice, most possible tasks likely related to a given one are usually created during the last few months. Consequently, what is the impact of using time filtering on the final recommendation results of the TaDeR system? Finally, how much does using deep learning features and the Siamese architecture help increase a TaDeR system's performance in chosen datasets? 

All experiments are operated on a computer with Intel(R)-Core(TM)-i7 2 CPUs running at 2.4GHz, 128GB of RAM, and an Nvidia GeForce RTX-2080Ti GPU. We present our experimental design and the corresponding results concerning these research questions in what follows. 

\subsection{Experimental Design}
As mentioned in the previous section, we first use dataset FLUME to find the best features and then compare all proposed methods on FLUME and MDLSITE. MDLSITE dataset has too many issues that took a long time to train for each model, so we only use FLUME for this experiment. We use five valuable attributes from the input data of a new Jira task created: Title (T), Description (D), Summary (S), Created Date, and Updated Date. For two specific Jira issues, we extract the following features: Title (T), Description (D), Summary (S), the gap between their created dates (CC or C2), the absolute difference between the created date of the current issue chosen and the updated date of another one, namely (CU).

In these experiments, we use four different numbers of top features: $K=1, 2, 3, 5$. For each value of K, we measure the performance of the corresponding type of features computed and the selected model using Accuracy@K, Recall@K and MRR@K. We also do extensive experiments by considering 28 combinations of five features computed (T, D, S, C2, and CU) to understand the impact of each type of feature combination of the model. These 28 features are as follows: T (Title), D (Description), S (Summary), TD (Title + Description), TS (Title + Summary), TDS (Title + Description + Summary), DS (Description + Summary), TC2 (Title +C2), DC2 (Description + C2), SC2 (Summary + C2),  TDC2 (Title + Description + C2), TSC2 (Title + Summary + C2), TDSC2 (Title + Description + Summary + C2), DSC2 (Description + Summary + C2), TCU (Title + CU), DCU (Description + CU), SCU (Summary + CU), TDCU (Title + Description + CU), TSCU (Title + Summary + CU), TDSCU (Title + Description + Summary + CU), DSCU (Description + Summary + CU), TC2CU (Title + C2 + CU), DC2CU (Description + C2 + CU), SC2CU (Summary + C2 + CU), TDC2CU (Title + Description + C2 + CU), TSC2CU (Title + Summary + C2 + CU), TDSC2CU (Title + Description + Summary+ C2 + CU), and DSC2CU (Description + Summary+ C2 + CU).

After finding out the best features, we tested these features with several combinations of two embedding methods (GloVe and FastText) and 3 model types (LSTM, GRU, and CNN).

Besides, we consider seven scenarios of applying the ``time filter'' to get the recommended items list in the post-processing step for a TaDeR system. In our work, we select the following ``time filter'': one month, two months, three months, and no time filter. 

For the traditional approach, we compute the corresponding performance of the proposed models (TF-IDF for feature extraction with several techniques for estimating matching scores) for datasets. 

\subsection{Hyperparameter Tuning}

After experiments, we began to tune our best model - GloVe Embedding and CNN model to get the best result.
We apply the hyperparameter tuning process to find our proposed Siamese architecture's best configuration for the TaDeR system. Those hyperparameters of all hidden layers (LSTM and Dense Layer) are the number of units and the action functions. Using two experimental datasets provided, we do the following hyperparameter tuning process:
\begin{itemize}
  \item The number of units can be finetuned as follows. All unit numbers of an LSTM layer and a fully connected (FC) layer start from 50 units at the beginning. Sequentially, we can increase each unit number by 50 units every step, and it will be stopped either the evaluation metrics stop growing or the loss value is not convergent.
  \item We have tried some typical activation functions (ReLU, LeakyReLU, and Sigmoid) for the Dense layer, and the activation function ReLU shows that it is the most sufficient in this problem.
\end{itemize}

\section{Discussions}

As mentioned in Section 6, we aimed to answer our first question about the attributes that are the best for our corresponding model. Since there are many issues in the MDLSITE dataset, we decided to perform the first 28 experiments on FLUME instead. We chose a combination of features in each experiment and applied them on GloVe and CNN model with the default 2-month timing filter. We have seven combinations with three textual features (Title – T, Description – D, Summary – S): T, D, S, DT, DS, DTS, and TS. Table \ref{tab:best-features-1} and Table \ref{tab:best-features-2} showed that DTS brings the best result among them with Accuracy@1 and MRR@1 reach 0.2026, Recall@1 obtains 0.1698. Using only title (T) and TS (title and summary) bring the worst results, with Accuracy@1 only 0.0529 and 0.0441.
To exploit the impact of our observation (issues created near each other usually have links between them), we added the feature CC, as mentioned in Section 5. After adding this feature, DTS + CC is the best result and is higher than DTS 0.01 in all Accuracy, MRR, and Recall.

The last updated time seemed to affect our problem, so we added the CU feature. DTS + CC + CU now is the best result in these experiments where MRR reaches 0.2291 and Recall can be obtained as 0.1956. This result is also higher than 0.01 compared with DTS + CC and 0.02 compared with DTS.
After knowing the best attributes for our corresponding model, we tried various embeddings and structures to have the best model.

In Table \ref{tab:experiment-none-1} and Table \ref{tab:experiment-none-2}, when we tried 6 different models: GloVe + LSTM, GloVe + GRU, GloVe + CNN, FastText + LSTM, FastText + GRU, and FastText + CNN with no timing filter on both dataset: FLUME and MDLSITE. In these experiments, these results are quite different between these datasets. With  FLUME, FastText + LSTM reaches the best result but in MDLSITE, FastText + GRU is the best model. However, these results are lower than cosine similarity and Euclidean distance in traditional approaches.

Adding a timing filter can speed up the computational time while still keeping the same as or better performance than searching in all historical Jira tasks of a given project.  From Table \ref{tab:experiment-1-month-1} to Table \ref{tab:experiment-3-month-2}, we tried a 1-month timing filter, 2-month timing filter and 3-month timing filter. With a 1-month timing filter, GloVe + CNN and FastText + LSTM bring the best results. Compared to FLUME, GloVe + CNN are the only models with Accuracy@1, MRR@1 higher than cosine similarity, and Euclidean distance in the traditional approach. Hence, we decided to propose this model. 

When analyzing the top 3 and top 5 results, Accuracy, MRR, and Recall were reduced when the timing filter increased from one month to three months, the same with the MDLSITE dataset. From this observation, a 1-month timing filter is the best timing filter for this problem.

We conclude from all analysis above that GloVe + CNN and a 1-month timing filter will be a proper model to solve our problem. This model reached 0.2335 in accuracy@1 and MRR@1 in the FLUME dataset. With the MDLSITE dataset, this model reaches 0.1258 in accuracy@1 and MRR@1. These results are higher than our best in the traditional approach, about 0.001 in accuracy@1 and MRR@1 and 0.2 in Recall@1 for MDLSITE. With FLUME, these results are higher than 0.001 in accuracy@1 and MRR@1 and 0.5 in Recall@1. With top 3 and top 5 on FLUME, accuracy@3 and Recall@3 are lower than the traditional approach, but MRR@3 is higher than 0.006, and MRR@5 and Recall@5 are higher 0.01 compared to the conventional methods.

\begin{table*}[]
\centering
\renewcommand{\arraystretch}{1.5}
\caption{Traditional methods on dataset FLUME - Top 1 and Top 2}
\label{tab:flume_traditional_1}
\scalebox{0.9}{
\begin{tabular}{|c|c|c|c|c|c|c|}
\hline
\multirow{2}{*}{\textbf{Distance   metrics}} & \multicolumn{3}{c|}{\textbf{Top 1}}                   & \multicolumn{3}{c|}{\textbf{Top 2}}                   \\ \cline{2-7} 
                                             & \textbf{Accuracy} & \textbf{MRR}    & \textbf{Recall} & \textbf{Accuracy} & \textbf{MRR}    & \textbf{Recall} \\ \hline
\textbf{Chebysev   Distance}                 & 0.1189            & 0.1189          & 0.0737          & 0.1189            & 0.1189          & 0.0769          \\ \hline
\textbf{Cosine   Similarity}                 & \textbf{0.2203}   & \textbf{0.2203} & \textbf{0.1589} & \textbf{0.2203}   & \textbf{0.2203} & \textbf{0.1589} \\ \hline
\textbf{Euclidean   Distance}                & \textbf{0.2203}   & \textbf{0.2203} & \textbf{0.1589} & \textbf{0.2203}   & \textbf{0.2203} & \textbf{0.1589} \\ \hline
\textbf{Manhattan   Distance}                & 0.1366            & 0.1366          & 0.0903          & 0.1366            & 0.1366          & 0.0903          \\ \hline
\end{tabular}}
\end{table*}

\begin{table*}[]
\centering
  \renewcommand{\arraystretch}{1.5}
\caption{Traditional methods on dataset FLUME - Top 3 and Top 5}
\label{tab:flume_traditional_2}
\scalebox{0.9}{
\begin{tabular}{|c|c|c|c|c|c|c|}
\hline
\multirow{2}{*}{\textbf{Distance   metrics}} & \multicolumn{3}{c|}{\textbf{Top 3}}                   & \multicolumn{3}{c|}{\textbf{Top 5}}                   \\ \cline{2-7} 
                                             & \textbf{Accuracy} & \textbf{MRR}    & \textbf{Recall} & \textbf{Accuracy} & \textbf{MRR}    & \textbf{Recall} \\ \hline
\textbf{Chebysev   Distance}                 & 0.1322            & 0.1233          & 0.1004          & 0.1454            & 0.1264          & 0.1153          \\ \hline
\textbf{Cosine   Similarity}                 & \textbf{0.2863}   & \textbf{0.2423} & \textbf{0.2392} & \textbf{0.3172}   & \textbf{0.2485} & \textbf{0.2584} \\ \hline
\textbf{Euclidean   Distance}                & \textbf{0.2863}   & \textbf{0.2423} & \textbf{0.2392} & \textbf{0.3172}   & \textbf{0.2485} & \textbf{0.2584} \\ \hline
\textbf{Manhattan   Distance}                & 0.1982            & 0.1571          & 0.1629          & 0.2291            & 0.1633          & 0.1903          \\ \hline
\end{tabular}}
\end{table*}

\begin{table*}[]
\centering
  \renewcommand{\arraystretch}{1.5}
\caption{Traditional methods on dataset MDLSITE - Top 1 and Top 2}
\label{tab:mdlsite_traditional_1}
\scalebox{0.9}{
\begin{tabular}{|c|c|c|c|c|c|c|}
\hline
\multirow{2}{*}{\textbf{Distance   metrics}} & \multicolumn{3}{c|}{\textbf{Top 1}}                   & \multicolumn{3}{c|}{\textbf{Top 2}}                   \\ \cline{2-7} 
                                             & \textbf{Accuracy} & \textbf{MRR}    & \textbf{Recall} & \textbf{Accuracy} & \textbf{MRR}    & \textbf{Recall} \\ \hline
\textbf{Chebysev   Distance}                 & 0.0679            & 0.0679          & 0.0454          & 0.0767            & 0.0723          & 0.0521          \\ \hline
\textbf{Cosine   Similarity}                 & \textbf{0.1245}   & \textbf{0.1245} & \textbf{0.0905} & \textbf{0.1245}   & \textbf{0.1245} & \textbf{0.0905} \\ \hline
\textbf{Euclidean   Distance}                & \textbf{0.1245}   & \textbf{0.1245} & \textbf{0.0905} & \textbf{0.1245}   & \textbf{0.1245} & \textbf{0.0905} \\ \hline
\textbf{Manhattan   Distance}                & 0.0893            & 0.0893          & 0.0632          & 0.0893            & 0.0893          & 0.0632          \\ \hline
\end{tabular}}
\end{table*}

\begin{table*}[ht!]
\centering
  \renewcommand{\arraystretch}{1.5}
\caption{Traditional methods on dataset MDLSITE - Top 3 and Top 5}
\label{tab:mdlsite_traditional_2}
\scalebox{0.9}{
\begin{tabular}{|c|c|c|c|c|c|c|}
\hline
\multirow{2}{*}{\textbf{Distance   metrics}} & \multicolumn{3}{c|}{\textbf{Top 3}}                   & \multicolumn{3}{c|}{\textbf{Top 5}}                   \\ \cline{2-7} 
                                             & \textbf{Accuracy} & \textbf{MRR}    & \textbf{Recall} & \textbf{Accuracy} & \textbf{MRR}    & \textbf{Recall} \\ \hline
\textbf{Chebysev   Distance}                 & 0.0855            & 0.0753          & 0.0595          & 0.0956            & 0.0773          & 0.0672          \\ \hline
\textbf{Cosine   Similarity}                 & \textbf{0.1434}   & \textbf{0.1308} & \textbf{0.1111} & \textbf{0.1547}   & \textbf{0.1331} & \textbf{0.1222} \\ \hline
\textbf{Euclidean   Distance}                & \textbf{0.1434}   & \textbf{0.1308} & \textbf{0.1111} & \textbf{0.1547}   & \textbf{0.1331} & \textbf{0.1222} \\ \hline
\textbf{Manhattan   Distance}                & 0.1019            & 0.0935          & 0.0766          & 0.1082            & 0.0948          & 0.0799          \\ \hline
\end{tabular}}
\end{table*}

\begin{table*}[]
\centering
  \renewcommand{\arraystretch}{1.5}
\caption{Best features experiment on FLUME - Top 1 and Top 2 (using the two-month timing filter)}
\label{tab:best-features-1}
\scalebox{0.9}{
\begin{tabular}{|l|c|c|c|c|c|c|}
\hline
\multicolumn{1}{|c|}{\multirow{2}{*}{\textbf{Features}}} & \multicolumn{3}{c|}{\textbf{Top 1}}                   & \multicolumn{3}{c|}{\textbf{Top 2}}                   \\ \cline{2-7} 
\multicolumn{1}{|c|}{}                                   & \textbf{Accuracy} & \textbf{MRR}    & \textbf{Recall} & \textbf{Accuracy} & \textbf{MRR}    & \textbf{Recall} \\ \hline
\textbf{T}                                               & 0.0529            & 0.0529          & 0.0395          & 0.0705            & 0.0771          & 0.0577          \\ \hline
\textbf{D}                                               & 0.1057            & 0.1057          & 0.0879          & 0.1189            & 0.1454          & 0.0982          \\ \hline
\textbf{S}                                               & 0.1145            & 0.1145          & 0.0905          & 0.1542            & 0.1674          & 0.1278          \\ \hline
\textbf{DT}                                              & 0.0749            & 0.0749          & 0.0624          & 0.1013            & 0.1145          & 0.0777          \\ \hline
\textbf{DS}                                              & 0.0749            & 0.0749          & 0.0628          & 0.0793            & 0.1079          & 0.0662          \\ \hline
\textbf{DTS}                                             & \textbf{0.2026}   & \textbf{0.2026} & \textbf{0.1698} & \textbf{0.2203}   & \textbf{0.2952} & \textbf{0.1824} \\ \hline
\textbf{TS}                                              & 0.0441            & 0.0441          & 0.0279          & 0.0749            & 0.0727          & 0.0558          \\ \hline
\textbf{T + CC}                                          & 0.0749            & 0.0749          & 0.0598          & 0.1233            & 0.1189          & 0.0994          \\ \hline
\textbf{D + CC}                                          & 0.1938            & 0.1938          & 0.1637          & 0.2115            & 0.2885          & 0.1809          \\ \hline
\textbf{S + CC}                                          & 0.0485            & 0.0485          & 0.0292          & 0.0837            & 0.0771          & 0.0615          \\ \hline
\textbf{DT + CC}                                         & 0.0969            & 0.0969          & 0.0770          & 0.1145            & 0.1366          & 0.0887          \\ \hline
\textbf{DS + CC}                                         & 0.0881            & 0.0881          & 0.0624          & 0.1101            & 0.1322          & 0.0862          \\ \hline
\textbf{DTS + CC}                                        & 0.2115            & 0.2115          & 0.1787          & 0.2247            & 0.3062          & 0.1876          \\ \hline
\textbf{TS + CC}                                         & 0.0485            & 0.0485          & 0.0406          & 0.0925            & 0.0859          & 0.0718          \\ \hline
\textbf{T + CU}                                          & 0.0705            & 0.0705          & 0.0527          & 0.1057            & 0.1167          & 0.0789          \\ \hline
\textbf{D + CU}                                          & 0.1718            & 0.1718          & 0.1559          & 0.2026            & 0.2555          & 0.1787          \\ \hline
\textbf{S + CU}                                          & 0.0573            & 0.0573          & 0.0492          & 0.1057            & 0.0947          & 0.0807          \\ \hline
\textbf{DT + CU}                                         & 0.0705            & 0.0705          & 0.0507          & 0.0881            & 0.1057          & 0.0665          \\ \hline
\textbf{DS + CU}                                         & 0.0749            & 0.0749          & 0.0606          & 0.0881            & 0.1101          & 0.0679          \\ \hline
\textbf{DTS + CU}                                        & 0.1762            & 0.1762          & 0.1545          & 0.2026            & 0.2709          & 0.1754          \\ \hline
\textbf{TS + CU}                                         & 0.1762            & 0.1762          & 0.1534          & 0.2291            & 0.2467          & 0.2030          \\ \hline
\textbf{T + CC + CU}                                     & 0.0617            & 0.0617          & 0.0499          & 0.0793            & 0.0903          & 0.0638          \\ \hline
\textbf{D + CC + CU}                                     & 0.1806            & 0.1806          & 0.1545          & 0.2115            & 0.2797          & 0.1838          \\ \hline
\textbf{S + CC + CU}                                     & 0.0793            & 0.0793          & 0.0564          & 0.1101            & 0.1057          & 0.0795          \\ \hline
\textbf{DT + CC + CU}                                    & 0.0793            & 0.0793          & 0.0613          & 0.1013            & 0.1189          & 0.0792          \\ \hline
\textbf{DS + CC + CU}                                    & 0.0705            & 0.0705          & 0.0562          & 0.0749            & 0.1013          & 0.0614          \\ \hline
\textbf{DTS + CC + CU}                                   & \textbf{0.2291}   & \textbf{0.2291} & \textbf{0.1956} & \textbf{0.2467}   & \textbf{0.3348} & \textbf{0.2095} \\ \hline
\textbf{TS + CC + CU}                                    & 0.0705            & 0.0705          & 0.0437          & 0.0925            & 0.1057          & 0.0645          \\ \hline
\end{tabular}}
\end{table*}

\begin{table*}[]
\centering
  \renewcommand{\arraystretch}{1.5}
\caption{Best features experiment on FLUME - Top 3 and Top 5  (using the two-month timing filter)}
\label{tab:best-features-2}
\scalebox{0.9}{
\begin{tabular}{|l|c|c|c|c|c|c|}
\hline
\multicolumn{1}{|c|}{\multirow{2}{*}{\textbf{Features}}} & \multicolumn{3}{c|}{\textbf{Top 3}}                   & \multicolumn{3}{c|}{\textbf{Top 5}}                   \\ \cline{2-7} 
\multicolumn{1}{|c|}{}                                   & \textbf{Accuracy} & \textbf{MRR}    & \textbf{Recall} & \textbf{Accuracy} & \textbf{MRR}    & \textbf{Recall} \\ \hline
\textbf{T}                                               & 0.1233            & 0.1035          & 0.0952          & 0.1630            & 0.1194          & 0.1241          \\ \hline
\textbf{D}                                               & 0.1454            & 0.1630          & 0.1177          & 0.1674            & 0.1883          & 0.1433          \\ \hline
\textbf{S}                                               & 0.2159            & 0.1982          & 0.1858          & 0.2996            & 0.2313          & 0.2618          \\ \hline
\textbf{DT}                                              & 0.1189            & 0.1322          & 0.0970          & 0.2203            & 0.1742          & 0.1819          \\ \hline
\textbf{DS}                                              & 0.1057            & 0.1211          & 0.0841          & 0.1806            & 0.1511          & 0.1549          \\ \hline
\textbf{DTS}                                             & \textbf{0.2511}   & \textbf{0.3172} & \textbf{0.2067} & \textbf{0.2731}   & \textbf{0.3423} & \textbf{0.2297} \\ \hline
\textbf{TS}                                              & 0.1101            & 0.0918          & 0.0777          & 0.1498            & 0.1123          & 0.1059          \\ \hline
\textbf{T + CC}                                          & 0.1718            & 0.1439          & 0.1446          & 0.2731            & 0.1785          & 0.2465          \\ \hline
\textbf{D + CC}                                          & 0.2247            & 0.3018          & 0.1934          & 0.2511            & 0.3262          & 0.2224          \\ \hline
\textbf{S + CC}                                          & 0.1322            & 0.0977          & 0.1044          & 0.2247            & 0.1338          & 0.1798          \\ \hline
\textbf{DT + CC}                                         & 0.1233            & 0.1468          & 0.0943          & 0.1278            & 0.1609          & 0.0985          \\ \hline
\textbf{DS + CC}                                         & 0.1586            & 0.1571          & 0.1298          & 0.2203            & 0.1838          & 0.1899          \\ \hline
\textbf{DTS + CC}                                        & 0.2467            & 0.3282          & 0.2032          & 0.2907            & 0.3498          & 0.2455          \\ \hline
\textbf{TS + CC}                                         & 0.1057            & 0.0977          & 0.0835          & 0.1410            & 0.1153          & 0.1073          \\ \hline
\textbf{T + CU}                                          & 0.1322            & 0.1329          & 0.0991          & 0.1806            & 0.1521          & 0.1456          \\ \hline
\textbf{D + CU}                                          & 0.2379            & 0.2878          & 0.2054          & 0.2643            & 0.3019          & 0.2297          \\ \hline
\textbf{S + CU}                                          & 0.1586            & 0.1182          & 0.1233          & 0.2731            & 0.1640          & 0.2234          \\ \hline
\textbf{DT + CU}                                         & 0.1278            & 0.1248          & 0.0995          & 0.2247            & 0.1662          & 0.1839          \\ \hline
\textbf{DS + CU}                                         & 0.1145            & 0.1248          & 0.0866          & 0.1586            & 0.1455          & 0.1243          \\ \hline
\textbf{DTS + CU}                                        & 0.2291            & 0.2915          & 0.1985          & 0.2467            & 0.3084          & 0.2105          \\ \hline
\textbf{TS + CU}                                         & 0.2555            & 0.2805          & 0.2213          & 0.2687            & 0.3045          & 0.2311          \\ \hline
\textbf{T + CC + CU}                                     & 0.1233            & 0.1094          & 0.0935          & 0.1762            & 0.1321          & 0.1385          \\ \hline
\textbf{D + CC + CU}                                     & 0.2203            & 0.2930          & 0.1929          & 0.2555            & 0.3145          & 0.2319          \\ \hline
\textbf{S + CC + CU}                                     & 0.1454            & 0.1351          & 0.1089          & 0.1762            & 0.1547          & 0.1408          \\ \hline
\textbf{DT + CC + CU}                                    & 0.1278            & 0.1395          & 0.1022          & 0.1938            & 0.1624          & 0.1664          \\ \hline
\textbf{DS + CC + CU}                                    & 0.0925            & 0.1116          & 0.0734          & 0.1189            & 0.1255          & 0.0891          \\ \hline
\textbf{DTS + CC + CU}                                   & \textbf{0.2555}   & \textbf{0.3495} & \textbf{0.2201} & \textbf{0.2819}   & \textbf{0.3728} & \textbf{0.2455} \\ \hline
\textbf{TS + CC + CU}                                    & 0.1410            & 0.1322          & 0.1118          & 0.2070            & 0.1608          & 0.1712          \\ \hline
\end{tabular}}
\end{table*}

\begin{table*}[]
\centering
  \renewcommand{\arraystretch}{1.5}
\caption{Experiments with no time filter - Top 1 and Top 2}
\label{tab:experiment-none-1}
\scalebox{0.9}{
\begin{tabular}{|c|c|c|c|c|c|c|}
\hline
\multirow{2}{*}{\textbf{Model}} & \multicolumn{3}{c|}{\textbf{Top 1}}                   & \multicolumn{3}{c|}{\textbf{Top 2}}                   \\ \cline{2-7} 
                                & \textbf{Accuracy} & \textbf{MRR}    & \textbf{Recall} & \textbf{Accuracy} & \textbf{MRR}    & \textbf{Recall} \\ \hline
\multicolumn{7}{|c|}{\textbf{FLUME}}                                                                                                            \\ \hline
\textbf{GloVe + LSTM}           & \textbf{0.0881}   & \textbf{0.0881} & \textbf{0.0630} & \textbf{0.2159}   & \textbf{0.1520} & \textbf{0.1774} \\ \hline
GloVe + GRU                     & 0.0441            & 0.0441          & 0.0183          & 0.1542            & 0.0991          & 0.1100          \\ \hline
GloVe + CNN                     & 0.0000            & 0.0000          & 0.0000          & 0.0000            & 0.0000          & 0.0000          \\ \hline
\textbf{FastText +   LSTM}      & \textbf{0.1674}   & \textbf{0.1674} & \textbf{0.1589} & \textbf{0.1894}   & \textbf{0.1784} & \textbf{0.1750} \\ \hline
FastText + GRU                  & 0.0881            & 0.0881          & 0.0837          & 0.1057            & 0.0969          & 0.0977          \\ \hline
FastText + CNN                  & 0.1233            & 0.1233          & 0.1149          & 0.1850            & 0.1542          & 0.1685          \\ \hline
\multicolumn{7}{|c|}{\textbf{MDLSITE}}                                                                                                          \\ \hline
\textbf{GloVe + LSTM}           & \textbf{0.0252}   & \textbf{0.0252} & \textbf{0.0160} & \textbf{0.0377}   & \textbf{0.0314} & \textbf{0.0260} \\ \hline
GloVe + GRU                     & 0.0000            & 0.0000          & 0.0000          & 0.0000            & 0.0000          & 0.0000          \\ \hline
GloVe + CNN                     & 0.0000            & 0.0000          & 0.0000          & 0.0000            & 0.0000          & 0.0000          \\ \hline
FastText +   LSTM               & 0.0013            & 0.0013          & 0.0001          & 0.2063            & 0.1038          & 0.1991          \\ \hline
\textbf{FastText + GRU}         & \textbf{0.0956}   & \textbf{0.0956} & \textbf{0.0848} & \textbf{0.2465}   & \textbf{0.1711} & \textbf{0.2351} \\ \hline
FastText + CNN                  & 0.0000            & 0.0000          & 0.0000          & 0.1447            & 0.0723          & 0.1432          \\ \hline
\end{tabular}}
\end{table*}

\begin{table*}[]
\centering
  \renewcommand{\arraystretch}{1.5}
\caption{Experiments with no time filter - Top 3 and Top 5}
\label{tab:experiment-none-2}
\scalebox{0.9}{
\begin{tabular}{|c|c|c|c|c|c|c|}
\hline
\multirow{2}{*}{\textbf{Model}} & \multicolumn{3}{c|}{\textbf{Top 3}}                   & \multicolumn{3}{c|}{\textbf{Top 5}}                   \\ \cline{2-7} 
                                & \textbf{Accuracy} & \textbf{MRR}    & \textbf{Recall} & \textbf{Accuracy} & \textbf{MRR}    & \textbf{Recall} \\ \hline
\multicolumn{7}{|c|}{\textbf{FLUME}}                                                                                                            \\ \hline
\textbf{GloVe + LSTM}           & \textbf{0.2467}   & \textbf{0.1623} & \textbf{0.2005} & \textbf{0.3392}   & \textbf{0.1827} & \textbf{0.2856} \\ \hline
GloVe + GRU                     & 0.2291            & 0.1241          & 0.1845          & 0.3040            & 0.1408          & 0.2515          \\ \hline
GloVe + CNN                     & 0.0000            & 0.0000          & 0.0000          & 0.0705            & 0.0170          & 0.0683          \\ \hline
\textbf{FastText +   LSTM}      & \textbf{0.2467}   & \textbf{0.1975} & \textbf{0.2326} & \textbf{0.3084}   & \textbf{0.2123} & \textbf{0.2916} \\ \hline
FastText + GRU                  & 0.1938            & 0.1263          & 0.1721          & 0.2379            & 0.1364          & 0.2101          \\ \hline
FastText + CNN                  & 0.1938            & 0.1571          & 0.1706          & 0.2159            & 0.1624          & 0.1909          \\ \hline
\multicolumn{7}{|c|}{\textbf{MDLSITE}}                                                                                                          \\ \hline
\textbf{GloVe + LSTM}           & \textbf{0.0503}   & \textbf{0.0356} & \textbf{0.0355} & \textbf{0.0755}   & \textbf{0.0416} & \textbf{0.0570} \\ \hline
GloVe + GRU                     & 0.0239            & 0.0080          & 0.0153          & 0.0478            & 0.0132          & 0.0337          \\ \hline
GloVe + CNN                     & 0.0252            & 0.0084          & 0.0252          & 0.0830            & 0.0206          & 0.0774          \\ \hline
FastText +   LSTM               & 0.3321            & 0.1457          & 0.3218          & 0.4453            & 0.1708          & 0.4311          \\ \hline
\textbf{FastText + GRU}         & \textbf{0.3799}   & \textbf{0.2155} & \textbf{0.3660} & \textbf{0.4969}   & \textbf{0.2427} & \textbf{0.4774} \\ \hline
FastText + CNN                  & 0.2818            & 0.1180          & 0.2779          & 0.4604            & 0.1612          & 0.4547          \\ \hline
\end{tabular}}
\end{table*}

\begin{table*}[]
\centering
  \renewcommand{\arraystretch}{1.5}
\caption{Experiments with time filter  1 month - Top 1 and Top 2}
\label{tab:experiment-1-month-1}
\scalebox{0.9}{
\begin{tabular}{|c|c|c|c|c|c|c|}
\hline
\multirow{2}{*}{\textbf{Model}} & \multicolumn{3}{c|}{\textbf{Top 1}}                   & \multicolumn{3}{c|}{\textbf{Top 2}}                   \\ \cline{2-7} 
                                & \textbf{Accuracy} & \textbf{MRR}    & \textbf{Recall} & \textbf{Accuracy} & \textbf{MRR}    & \textbf{Recall} \\ \hline
\multicolumn{7}{|c|}{\textbf{FLUME}}                                                                                                            \\ \hline
GloVe + LSTM                    & 0.0837            & 0.0837          & 0.0586          & 0.2115            & 0.1476          & 0.1730          \\ \hline
GloVe + GRU                     & 0.0396            & 0.0396          & 0.0139          & 0.1498            & 0.0947          & 0.1056          \\ \hline
\textbf{GloVe + CNN}            & \textbf{0.2335}   & \textbf{0.2335} & \textbf{0.2011} & \textbf{0.2511}   & \textbf{0.2423} & \textbf{0.2216} \\ \hline
\textbf{FastText +   LSTM}      & \textbf{0.2026}   & \textbf{0.2026} & \textbf{0.1882} & \textbf{0.2335}   & \textbf{0.2181} & \textbf{0.2144} \\ \hline
FastText + GRU                  & 0.1410            & 0.1410          & 0.1210          & 0.1674            & 0.1542          & 0.1411          \\ \hline
FastText + CNN                  & 0.2026            & 0.2026          & 0.1627          & 0.2070            & 0.2048          & 0.1699          \\ \hline
\multicolumn{7}{|c|}{\textbf{MDLSITE}}                                                                                                          \\ \hline
GloVe + LSTM                    & 0.0252            & 0.0252          & 0.0160          & 0.0377            & 0.0314          & 0.0260          \\ \hline
GloVe + GRU                     & 0.0314            & 0.0314          & 0.0211          & 0.0415            & 0.0365          & 0.0290          \\ \hline
\textbf{GloVe + CNN}            & \textbf{0.1258}   & \textbf{0.1258} & \textbf{0.1141} & \textbf{0.2541}   & \textbf{0.1899} & \textbf{0.2401} \\ \hline
\textbf{FastText +   LSTM}      & \textbf{0.2302}   & \textbf{0.2302} & \textbf{0.2243} & \textbf{0.3610}   & \textbf{0.2956} & \textbf{0.3537} \\ \hline
FastText + GRU                  & 0.1308            & 0.1308          & 0.1216          & 0.2881            & 0.2094          & 0.2762          \\ \hline
FastText + CNN                  & 0.1836            & 0.1836          & 0.1779          & 0.3233            & 0.2535          & 0.3136          \\ \hline
\end{tabular}}
\end{table*}

\begin{table*}[]
\centering
  \renewcommand{\arraystretch}{1.5}
\caption{Experiments with time filter  1 month - Top 3 and Top 5}
\label{tab:experiment-1-month-2}
\scalebox{0.9}{
\begin{tabular}{|c|c|c|c|c|c|c|}
\hline
\multirow{2}{*}{\textbf{Model}} & \multicolumn{3}{c|}{\textbf{Top 3}}                   & \multicolumn{3}{c|}{\textbf{Top 5}}                   \\ \cline{2-7} 
                                & \textbf{Accuracy} & \textbf{MRR}    & \textbf{Recall} & \textbf{Accuracy} & \textbf{MRR}    & \textbf{Recall} \\ \hline
\multicolumn{7}{|c|}{\textbf{FLUME}}                                                                                                            \\ \hline
GloVe + LSTM                    & 0.2379            & 0.1564          & 0.1917          & 0.3304            & 0.1769          & 0.2767          \\ \hline
GloVe + GRU                     & 0.2203            & 0.1182          & 0.1757          & 0.2952            & 0.1349          & 0.2426          \\ \hline
\textbf{GloVe + CNN}            & \textbf{0.2687}   & \textbf{0.2482} & \textbf{0.2377} & \textbf{0.3040}   & \textbf{0.2563} & \textbf{0.2651} \\ \hline
\textbf{FastText +   LSTM}      & \textbf{0.2996}   & \textbf{0.2401} & \textbf{0.2786} & \textbf{0.3524}   & \textbf{0.2520} & \textbf{0.3225} \\ \hline
FastText + GRU                  & 0.2555            & 0.1836          & 0.2225          & 0.3040            & 0.1943          & 0.2698          \\ \hline
FastText + CNN                  & 0.2247            & 0.2107          & 0.1890          & 0.2907            & 0.2253          & 0.2535          \\ \hline
\multicolumn{7}{|c|}{\textbf{MDLSITE}}                                                                                                          \\ \hline
GloVe + LSTM                    & 0.0503            & 0.0356          & 0.0355          & 0.0767            & 0.0418          & 0.0576          \\ \hline
GloVe + GRU                     & 0.0516            & 0.0398          & 0.0371          & 0.0780            & 0.0461          & 0.0581          \\ \hline
\textbf{GloVe + CNN}            & \textbf{0.3975}   & \textbf{0.2377} & \textbf{0.3765} & \textbf{0.5270}   & \textbf{0.2689} & \textbf{0.5018} \\ \hline
\textbf{FastText +   LSTM}      & \textbf{0.4277}   & \textbf{0.3178} & \textbf{0.4157} & \textbf{0.5220}   & \textbf{0.3386} & \textbf{0.5053} \\ \hline
FastText + GRU                  & 0.3925            & 0.2442          & 0.3792          & 0.5308            & 0.2753          & 0.5103          \\ \hline
FastText + CNN                  & 0.4730            & 0.3034          & 0.4594          & 0.5648            & 0.3230          & 0.5418          \\ \hline
\end{tabular}}
\end{table*}

\begin{table*}[ht!]
\centering
  \renewcommand{\arraystretch}{1.5}
\caption{Experiments with time filter 2 months - Top 1 and Top 2}
\label{tab:experiment-2-month-1}
\scalebox{0.9}{
\begin{tabular}{|c|c|c|c|c|c|c|}
\hline
\multirow{2}{*}{\textbf{Model}} & \multicolumn{3}{c|}{\textbf{Top 1}}                   & \multicolumn{3}{c|}{\textbf{Top 2}}                   \\ \cline{2-7} 
                                & \textbf{Accuracy} & \textbf{MRR}    & \textbf{Recall} & \textbf{Accuracy} & \textbf{MRR}    & \textbf{Recall} \\ \hline
\multicolumn{7}{|c|}{\textbf{FLUME}}                                                                                                            \\ \hline
GloVe + LSTM                    & 0.0837            & 0.0837          & 0.0586          & 0.2115            & 0.1476          & 0.1730          \\ \hline
GloVe + GRU                     & 0.0396            & 0.0396          & 0.0139          & 0.1498            & 0.0947          & 0.1056          \\ \hline
\textbf{GloVe + CNN}            & \textbf{0.2291}   & \textbf{0.2291} & \textbf{0.1956} & \textbf{0.2467}   & \textbf{0.2379} & \textbf{0.2095} \\ \hline
FastText +   LSTM               & 0.1718            & 0.1718          & 0.1633          & 0.1982            & 0.1850          & 0.1846          \\ \hline
FastText + GRU                  & 0.1013            & 0.1013          & 0.0914          & 0.1366            & 0.1189          & 0.1159          \\ \hline
\textbf{FastText + CNN}         & \textbf{0.1762}   & \textbf{0.1762} & \textbf{0.1523} & \textbf{0.2026}   & \textbf{0.1894} & \textbf{0.1636} \\ \hline
\multicolumn{7}{|c|}{\textbf{MDLSITE}}                                                                                                          \\ \hline
GloVe + LSTM                    & 0.0252            & 0.0252          & 0.0160          & 0.0377            & 0.0314          & 0.0260          \\ \hline
GloVe + GRU                     & 0.0289            & 0.0289          & 0.0185          & 0.0390            & 0.0340          & 0.0265          \\ \hline
\textbf{GloVe + CNN}            & \textbf{0.1346}   & \textbf{0.1346} & \textbf{0.1165} & \textbf{0.2667}   & \textbf{0.2006} & \textbf{0.2461} \\ \hline
\textbf{FastText +   LSTM}      & \textbf{0.2289}   & \textbf{0.2289} & \textbf{0.2212} & \textbf{0.3560}   & \textbf{0.2925} & \textbf{0.3477} \\ \hline
FastText + GRU                  & 0.0981            & 0.0981          & 0.0890          & 0.2792            & 0.1887          & 0.2657          \\ \hline
FastText + CNN                  & 0.1811            & 0.1811          & 0.1757          & 0.3157            & 0.2484          & 0.3064          \\ \hline
\end{tabular}}
\end{table*}

\begin{table*}[]
\centering
  \renewcommand{\arraystretch}{1.5}
\caption{Experiments with time filter 2 months - Top 3 and Top 5}
\label{tab:experiment-2-month-2}
\scalebox{0.9}{
\begin{tabular}{|c|c|c|c|c|c|c|}
\hline
\multirow{2}{*}{\textbf{Model}} & \multicolumn{3}{c|}{\textbf{Top 3}}                   & \multicolumn{3}{c|}{\textbf{Top 5}}                   \\ \cline{2-7} 
                                & \textbf{Accuracy} & \textbf{MRR}    & \textbf{Recall} & \textbf{Accuracy} & \textbf{MRR}    & \textbf{Recall} \\ \hline
\multicolumn{7}{|c|}{\textbf{FLUME}}                                                                                                            \\ \hline
GloVe + LSTM                    & 0.2379            & 0.1564          & 0.1917          & 0.3304            & 0.1769          & 0.2767          \\ \hline
GloVe + GRU                     & 0.2203            & 0.1182          & 0.1757          & 0.2952            & 0.1349          & 0.2426          \\ \hline
\textbf{GloVe + CNN}            & \textbf{0.2555}   & \textbf{0.2408} & \textbf{0.2201} & \textbf{0.2819}   & \textbf{0.2470} & \textbf{0.2455} \\ \hline
FastText +   LSTM               & 0.2687            & 0.2085          & 0.2553          & 0.3128            & 0.2189          & 0.2910          \\ \hline
FastText + GRU                  & 0.2115            & 0.1439          & 0.1834          & 0.2467            & 0.1516          & 0.2164          \\ \hline
\textbf{FastText + CNN}         & \textbf{0.2247}   & \textbf{0.1968} & \textbf{0.1860} & \textbf{0.2555}   & \textbf{0.2034} & \textbf{0.2190} \\ \hline
\multicolumn{7}{|c|}{\textbf{MDLSITE}}                                                                                                          \\ \hline
GloVe + LSTM                    & 0.0503            & 0.0356          & 0.0355          & 0.0755            & 0.0416          & 0.0570          \\ \hline
GloVe + GRU                     & 0.0516            & 0.0382          & 0.0371          & 0.0780            & 0.0444          & 0.0581          \\ \hline
\textbf{GloVe + CNN}            & \textbf{0.3786}   & \textbf{0.2379} & \textbf{0.3553} & \textbf{0.5182}   & \textbf{0.2701} & \textbf{0.4929} \\ \hline
\textbf{FastText +   LSTM}      & \textbf{0.4226}   & \textbf{0.3147} & \textbf{0.4105} & \textbf{0.5145}   & \textbf{0.3350} & \textbf{0.4981} \\ \hline
FastText + GRU                  & 0.3824            & 0.2231          & 0.3662          & 0.5182            & 0.2546          & 0.4957          \\ \hline
FastText + CNN                  & 0.4642            & 0.2979          & 0.4499          & 0.5258            & 0.3117          & 0.5055          \\ \hline
\end{tabular}}
\end{table*}

\begin{table*}[ht!]
\centering
  \renewcommand{\arraystretch}{1.5}
\caption{Experiments with time filter 3 months - Top 1 and Top 2}
\label{tab:experiment-3-month-1}
\scalebox{0.9}{
\begin{tabular}{|c|c|c|c|c|c|c|}
\hline
\multirow{2}{*}{\textbf{Model}} & \multicolumn{3}{c|}{\textbf{Top 1}}                   & \multicolumn{3}{c|}{\textbf{Top 2}}                   \\ \cline{2-7} 
                                & \textbf{Accuracy} & \textbf{MRR}    & \textbf{Recall} & \textbf{Accuracy} & \textbf{MRR}    & \textbf{Recall} \\ \hline
\multicolumn{7}{|c|}{\textbf{FLUME}}                                                                                                            \\ \hline
GloVe + LSTM                    & 0.0837            & 0.0837          & 0.0586          & 0.2115            & 0.1476          & 0.1730          \\ \hline
GloVe + GRU                     & 0.0396            & 0.0396          & 0.0139          & 0.1498            & 0.0947          & 0.1056          \\ \hline
\textbf{GloVe + CNN}            & \textbf{0.2247}   & \textbf{0.2247} & \textbf{0.1986} & \textbf{0.2423}   & \textbf{0.2335} & \textbf{0.2092} \\ \hline
FastText +   LSTM               & 0.1674            & 0.1674          & 0.1589          & 0.1982            & 0.1828          & 0.1846          \\ \hline
FastText + GRU                  & 0.1013            & 0.1013          & 0.0914          & 0.1322            & 0.1167          & 0.1153          \\ \hline
\textbf{FastText + CNN}         & \textbf{0.1718}   & \textbf{0.1718} & \textbf{0.1512} & \textbf{0.1894}   & \textbf{0.1806} & \textbf{0.1592} \\ \hline
\multicolumn{7}{|c|}{\textbf{MDLSITE}}                                                                                                          \\ \hline
GloVe + LSTM                    & 0.0252            & 0.0252          & 0.0160          & 0.0377            & 0.0314          & 0.0260          \\ \hline
GloVe + GRU                     & 0.0264            & 0.0264          & 0.0158          & 0.0377            & 0.0321          & 0.0250          \\ \hline
GloVe + CNN                     & 0.1258            & 0.1258          & 0.1049          & 0.2679            & 0.1969          & 0.2487          \\ \hline
\textbf{FastText +   LSTM}      & \textbf{0.2277}   & \textbf{0.2277} & \textbf{0.2199} & \textbf{0.3346}   & \textbf{0.2811} & \textbf{0.3249} \\ \hline
FastText + GRU                  & 0.0931            & 0.0931          & 0.0846          & 0.2465            & 0.1698          & 0.2340          \\ \hline
\textbf{FastText + CNN}         & \textbf{0.1774}   & \textbf{0.1774} & \textbf{0.1734} & \textbf{0.3157}   & \textbf{0.2465} & \textbf{0.3060} \\ \hline
\end{tabular}}
\end{table*}

\begin{table*}[]
\centering
  \renewcommand{\arraystretch}{1.5}
\caption{Experiments with time filter 3 months - Top 3 and Top 5}
\label{tab:experiment-3-month-2}
\scalebox{0.9}{
\begin{tabular}{|c|c|c|c|c|c|c|}
\hline
\multirow{2}{*}{\textbf{Model}} & \multicolumn{3}{c|}{\textbf{Top 3}}                   & \multicolumn{3}{c|}{\textbf{Top 5}}                   \\ \cline{2-7} 
                                & \textbf{Accuracy} & \textbf{MRR}    & \textbf{Recall} & \textbf{Accuracy} & \textbf{MRR}    & \textbf{Recall} \\ \hline
\multicolumn{7}{|c|}{\textbf{FLUME}}                                                                                                            \\ \hline
GloVe + LSTM                    & 0.2379            & 0.1564          & 0.1917          & 0.3304            & 0.1769          & 0.2767          \\ \hline
GloVe + GRU                     & 0.2203            & 0.1182          & 0.1757          & 0.2952            & 0.1349          & 0.2426          \\ \hline
\textbf{GloVe + CNN}            & \textbf{0.2467}   & \textbf{0.2349} & \textbf{0.2127} & \textbf{0.2775}   & \textbf{0.2420} & \textbf{0.2396} \\ \hline
FastText +   LSTM               & 0.2643            & 0.2048          & 0.2509          & 0.3084            & 0.2150          & 0.2920          \\ \hline
FastText + GRU                  & 0.2115            & 0.1432          & 0.1826          & 0.2379            & 0.1491          & 0.2101          \\ \hline
\textbf{FastText + CNN}         & \textbf{0.2115}   & \textbf{0.1880} & \textbf{0.1787} & \textbf{0.2423}   & \textbf{0.1946} & \textbf{0.2061} \\ \hline
\multicolumn{7}{|c|}{\textbf{MDLSITE}}                                                                                                          \\ \hline
GloVe + LSTM                    & 0.0503            & 0.0356          & 0.0355          & 0.0755            & 0.0416          & 0.0570          \\ \hline
GloVe + GRU                     & 0.0528            & 0.0371          & 0.0375          & 0.0767            & 0.0426          & 0.0579          \\ \hline
GloVe + CNN                     & 0.3560            & 0.2262          & 0.3357          & 0.5195            & 0.2641          & 0.4943          \\ \hline
\textbf{FastText +   LSTM}      & \textbf{0.4063}   & \textbf{0.3050} & \textbf{0.3935} & \textbf{0.5094}   & \textbf{0.3284} & \textbf{0.4945} \\ \hline
FastText + GRU                  & 0.3824            & 0.2151          & 0.3660          & 0.5157            & 0.2462          & 0.4932          \\ \hline
\textbf{FastText + CNN}         & \textbf{0.4553}   & \textbf{0.2931} & \textbf{0.4444} & \textbf{0.5182}   & \textbf{0.3072} & \textbf{0.5000} \\ \hline
\end{tabular}}
\end{table*}

\begin{table}[ht!]
\centering
\renewcommand{\arraystretch}{1.5}
\caption{The list of features used in Jira and their definitions (Part 1)}
\label{tab:feature-description1}
\scalebox{0.9}{
\begin{tabular}{|l|l|}
\hline
\multicolumn{1}{|c|}{\textbf{Features}}  & \multicolumn{1}{c|}{\textbf{Feature Description}}                                                                                                                                                      \\ \hline
\textbf{Title}                           & Title of the issue                                                                                                                                                                                     \\ \hline
\textbf{Link}                            & Public link of the issue                                                                                                                                                                               \\ \hline
\textbf{Project}                         & Name of the project that the issue is belonged to                                                                                                                                                      \\ \hline
\textbf{Description}                     & Full representation of an issue                                                                                                                                                                        \\ \hline
\textbf{Environment}                     & Short representation about the issue’s environment                                                                                                                                                     \\ \hline
\textbf{Key}                             & A string to represent for the issue                                                                                                                                                                    \\ \hline
\textbf{Summary}                         & \begin{tabular}[c]{@{}l@{}}A short description of the issue. Some summaries are \\ similar to Title\end{tabular}                                                                                       \\ \hline
\textbf{Type}                            & Type of the issue                                                                                                                                                                                      \\ \hline
\textbf{Parent}                          & \begin{tabular}[c]{@{}l@{}}Parent of the issue (Only relevant if the issue is a \\ sub-task)\end{tabular}                                                                                              \\ \hline
\textbf{Priority}                        & Priority level of the issue                                                                                                                                                                            \\ \hline
\textbf{Status}                          & State of the issue                                                                                                                                                                                     \\ \hline
\textbf{StatusCategory}                  & Group of state that the issue is belonged to                                                                                                                                                           \\ \hline
\textbf{Resolution}                      & Specifies the reason an issue is closed                                                                                                                                                                \\ \hline
\textbf{Labels}                          & All labels of the issue. (mostly NaN)                                                                                                                                                                  \\ \hline
\textbf{Assignee}                        & The user who was assigned to the issue                                                                                                                                                                 \\ \hline
\textbf{Reporter}                        & The user who created the issue                                                                                                                                                                         \\ \hline
\textbf{Security}                        & \begin{tabular}[c]{@{}l@{}}Security level (Only relevant if a security level \\ has been applied to the issue)\end{tabular}                                                                            \\ \hline
\textbf{Created}                         & The lastest date the issue was updated                                                                                                                                                                 \\ \hline
\textbf{Updated}                         & The lastest date the issue was updated                                                                                                                                                                 \\ \hline
\textbf{Resolved}                        & The last time the resolution was updated                                                                                                                                                               \\ \hline
\textbf{Due}                             & Estimated deadline of the issue                                                                                                                                                                        \\ \hline
\textbf{Versions}                        & \begin{tabular}[c]{@{}l@{}}Details the versions of product that the issue \\ affects\end{tabular}                                                                                                      \\ \hline
\textbf{FixVersion}                      & \begin{tabular}[c]{@{}l@{}}The versions that the work on the issue are \\ released\end{tabular}                                                                                                        \\ \hline
\textbf{Component}                       & \begin{tabular}[c]{@{}l@{}}Jira project components are generic \\ containers for issues\end{tabular}                                                                                                   \\ \hline
\textbf{Votes}                           & Number of votes                                                                                                                                                                                        \\ \hline
\textbf{Comments}                        & List of comments                                                                                                                                                                                       \\ \hline
\textbf{Attachments}                     & \begin{tabular}[c]{@{}l@{}}Files included in the issue (Only available if \\ the administrator has enabled attachments)\end{tabular}                                                                   \\ \hline
\textbf{TimeOriginalEstimate}            & \begin{tabular}[c]{@{}l@{}}Time was originally thought it would take to \\ complete the issue (Only available if \\ administrator has enabled time tracking)\end{tabular}                              \\ \hline
\textbf{TimeEstimate}                    & \begin{tabular}[c]{@{}l@{}}Actual time estimated would take to complete\\  the issue (Only available if administrator has \\ enabled time tracking)\end{tabular}                                       \\ \hline

\end{tabular}}
\end{table}

\begin{table}[ht!]
\centering
\renewcommand{\arraystretch}{1.5}
\caption{The list of features used in Jira and their definitions (Part 2)}
\label{tab:feature-description2}
\scalebox{0.9}{
\begin{tabular}{|l|l|}
\hline
\multicolumn{1}{|c|}{\textbf{Features}}  & \multicolumn{1}{c|}{\textbf{Feature Description}}  
\\ \hline 
\textbf{TimeSpent}                       & \begin{tabular}[c]{@{}l@{}}Time has been spending to complete the issue \\ (Only available if administrator has enabled \\ time tracking)\end{tabular}                                                 \\ \hline
\textbf{AggregateTime-OriginalEstimate}  & \begin{tabular}[c]{@{}l@{}}Time was originally thought it would take to \\ complete the issue plus all its subtasks (Only \\ available if the administrator has enabled \\ time tracking)\end{tabular} \\ \hline
\textbf{AggregateTime-RemainingEstimate} & \begin{tabular}[c]{@{}l@{}}Actually time estimated would take to complete \\ the issueplus all its subtasks(Only available if \\ administrator has enabled time tracking)\end{tabular}                 \\ \hline                        
\textbf{AggregateTimeSpent}              & \begin{tabular}[c]{@{}l@{}}Time has been spending to complete the issue plus\\ all its subtasks (Only available if  administrator\\ has enabled  time tracking)\end{tabular}                          \\ \hline
\textbf{IssueLinks}                      & \begin{tabular}[c]{@{}l@{}}All links the issue has (Only relevant if the issue \\is linked with at least one other issue)\end{tabular}                                                                \\ \hline
\textbf{Subtasks}                        & \begin{tabular}[c]{@{}l@{}}subtask of the issue (Only relevant if the issue\\ has subtasks)\end{tabular}                                                                                              \\ \hline
\textbf{Customfields}                    & Custom fields which were added to the issue                                                                                                                                                            \\ \hline
\end{tabular}}
\end{table}

\begin{table}[]
\renewcommand{\arraystretch}{1.5}
\centering
\caption{The list of features used in Jira and examples}
\label{tab:feature-example}
\begin{tabular}{|l|l|}
\hline
\multicolumn{1}{|c|}{\textbf{Features}}  & \multicolumn{1}{c|}{\textbf{Example}}                                                                                \\ \hline
\textbf{Title}                           & \begin{tabular}[c]{@{}l@{}}Title of the issue \& System contains $40$ invalid \\ user accounts with ...\end{tabular} \\ \hline
\textbf{Link}                            & https:\/\/tracker.moodle.org\/browse\/MDLSITE-3                                                                      \\ \hline
\textbf{Project}                         & \begin{tabular}[c]{@{}l@{}}Name of the project that the issue is belonged \\ to \& MDLSITE\end{tabular}              \\ \hline
\textbf{Description}                     & This is an issue from conversion from old s...                                                                       \\ \hline
\textbf{Environment}                     & None                                                                                                                 \\ \hline
\textbf{Key}                             & MDLSITE-3                                                                                                            \\ \hline
\textbf{Summary}                         & \begin{tabular}[c]{@{}l@{}}System contains 40\% invalid user accounts \\with ...  \end{tabular}                                                                 \\ \hline
\textbf{Type}                            & Task                                                                                                                 \\ \hline
\textbf{Parent}                          & NaN                                                                                                                  \\ \hline
\textbf{Priority}                        & Task                                                                                                                 \\ \hline
\textbf{Status}                          & Task                                                                                                                 \\ \hline
\textbf{StatusCategory}                  & Task                                                                                                                 \\ \hline
\textbf{Resolution}                      & Fixed                                                                                                                \\ \hline
\textbf{Labels}                          & NaN                                                                                                                  \\ \hline
\textbf{Assignee}                        & Michael Blake                                                                                                        \\ \hline
\textbf{Reporter}                        & mblake                                                                                                               \\ \hline
\textbf{Security}                        & NaN                                                                                                                  \\ \hline
\textbf{Created}                         & 2006-08-23 01:21:46                                                                                                  \\ \hline
\textbf{Updated}                         & 2006-08-25 21:59:07                                                                                                  \\ \hline
\textbf{Resolved}                        & 2006-08-25 21:57:50                                                                                                  \\ \hline
\textbf{Due}                             & NaN                                                                                                                  \\ \hline
\textbf{Versions}                        & NaN                                                                                                                  \\ \hline
\textbf{FixVersion}                      & NaN                                                                                                                  \\ \hline
\textbf{Component}                       & tracker.moodle.org                                                                                                   \\ \hline
\textbf{Votes}                           & 0                                                                                                                    \\ \hline
\textbf{Comments}                        & [`Issues assigned to invalid users have bee...                                                                       \\ \hline
\textbf{Attachments}                     & [`None']                                                                                                             \\ \hline
\textbf{TimeOriginalEstimate}            & NaN                                                                                                                  \\ \hline
\textbf{TimeEstimate}                    & NaN                                                                                                                  \\ \hline
\textbf{TimeSpent}                       & NaN                                                                                                                  \\ \hline
\textbf{AggregateTime-OriginalEstimate}  & NaN                                                                                                                  \\ \hline
\textbf{AggregateTime-RemainingEstimate} & NaN                                                                                                                  \\ \hline
\textbf{AggregateTimeSpent}              & NaN                                                                                                                  \\ \hline
\textbf{IssueLinks}                      & NaN                                                                                                                  \\ \hline
\textbf{Subtasks}                        & NaN                                                                                                                  \\ \hline
\textbf{Customfields}                    & \begin{tabular}[c]{@{}l@{}}[`Component watchers', `DaysSinceLast\\ Comment',...]\end{tabular}                        \\ \hline
\end{tabular}
\end{table}

\section{Conclusion and Future Works}
We have presented an extensive study for building a task dependency recommendation system in project management platforms. We have compared two different methods by considering a traditional one (using TF-IDF features and many distance metrics for estimating the matching score between the current task created and a specific task) and proposing an efficient Siamese network (using one GloVe Embedding, one CNN layer, and a Dense layer) for building the corresponding recommendation. We compare these methods using datasets (MDLSITE and FLUME) and the performance metrics Recall@K, Accuracy@K, and MRR@K. The experimental results show that the proposed method can outperform the traditional one in all datasets (increasing about 0.01 in Accuracy and MRR and 0.05 in Recall). Also, using the time filter can efficiently help enhance the performance of the TaDeR system.

There are still some limitations related to the TaDeR system.  The current problem only focuses on recommending the top relevant tasks with a given Jira observation created. It is fascinating if the TaDeR system can suggest top-related tasks and classify users' corresponding types of links. We aim to apply other embedding methods and extend our experiments to other challenging datasets in future work.

\nocite{*}
\bibliographystyle{spbasic}      
\bibliography{reference}   

\end{document}


\begin{frontmatter}

\title{TaDeR: A New Task Dependency Recommendation for Project Management Platforms
}

\author[seventhaddress,eighthaddress,ninthaddresss]{Quynh Nguyen}
\author[seventhaddress,eighthaddress,ninthaddress]{Dac H. Nguyen}
\author[seventhaddress,eighthaddress,ninthaddress]{Son T. Huynh}
\author[fourthaddress]{Hoa Dam}
\author[seventhaddress,eighthaddress,ninthaddress]{Binh T. Nguyen\corref{correspondingauthor}}
\cortext[correspondingauthor]{Corresponding author}
\ead{ngtbinh@hcmus.edu.vn}
\address[fourthaddress]{University of Wollongong, Australia}
\address[seventhaddress]{AISIA Research Lab, Ho Chi Minh City, Vietnam}
\address[eighthaddress]{Department of Computer Science, University of Science, Ho Chi Minh City, Vietnam}
\address[ninthaddress]{Vietnam National University, Ho Chi Minh City, Vietnam}

\end{frontmatter}